\documentstyle[12pt]{article}

\begin{document}

\setcounter{page}{1}

\pagestyle{plain}
\vspace{1cm}
\begin{center}
\Large{\bf On the Existence of the Logarithmic Correction Term in
Black Hole Entropy-Area Relation}\\
\small \vspace{1cm}
{\bf Kourosh Nozari}\quad and \quad{\bf A. S.  Sefidgar}\\
\vspace{0.5cm} {\it Department of Physics,
Faculty of Basic Science,\\
University of Mazandaran,\\
P. O. Box 47416-1467,
Babolsar, IRAN\\
e-mail: knozari@umz.ac.ir}
\end{center}
\vspace{1.5cm}
\begin{abstract}
In this paper we consider a model universe with large extra
dimensions to obtain a modified black hole entropy-area relation. We
use the generalized uncertainty principle to find a relation between
the number of spacetime dimensions and the presence or vanishing of
logarithmic prefactor in the black hole entropy-area relation. Our
calculations are restricted to the microcanonical ensembles and we
show that in the modified entropy-area relation, the microcanonical
logarithmic prefactor appears only when spacetime
has an even number of dimensions.
\end{abstract}
\newpage
\section{Motivation}
In Bekenstein-Hawking formalism of the black hole thermodynamics,
entropy of the black hole is given by the following famous formula
\begin{equation}
S_{BH}=\frac{A}{4L_p^2}
\end{equation}
where $A$ is the cross-sectional area of the black hole horizon.
Here spacetime is restricted to be a four dimensional manifold and
the black hole is considered as a macroscopically large object.
Originally, black hole entropy-area relation (1), was followed by
some thermodynamical considerations. However, a statistical
interpretation of the issue should be taken into account. A thorough
statistical framework of the problem is provided by quantum gravity
considerations. Several alternative approaches such as string
theory, loop quantum gravity and noncommutative geometry have
successfully been applied to find quantum gravitational corrections
of black hole entropy-area relation. According to these approaches,
a quantum corrected entropy of the black hole which may be observer
independent, can be written as follows[1]
\begin{equation}
S_{BH}=\frac{A}{4L_p^2}+ c_0 \ln(\frac{A}{4L_p^2})+\sum_{n=1}^\infty
c_n (\frac{A}{4L_p^2})^{-n} + const.
\end{equation}
The values of $c_0$ and $c_{n}$ are quantum gravity model dependent.
$c_0$, which is called logarithmic prefactor, contains some
information about details of the underlying quantum gravity
proposal. A knowledge of its exact values in different approaches
may reflect possible links between alternative quantum gravity
scenarios[2]. Nevertheless, there are some controversies in existing
literatures regarding the values of this prefactor even in a given
scenario. Many authors have tried to determine the possible values
of $c_0$. The matter which is obvious from the beginning is the fact
that $c_{0}$ is an ensemble dependent quantity. Hod has employed
statistical argument that constrain this prefactor to be a
non-negative integer[3]. Medved has studied the quantum correction
of microcanonical entropy of a fixed energy black hole[4]. He also
has studied quantum corrections to the canonical entropy. He has
argued that microcanonical entropy calculation leads to a
logarithmic prefactor with nonzero value of
${c_0}_{mc}=-\frac{3}{2}$ and the canonical entropy gives the value
of ${c_0}_{c}=\frac{3}{2}$. As a result, the total logarithmic
prefactor will be zero due to mutual cancelation of microcanonical
and canonical contributions. This conclusion is consistent with
Hod's result since it gives a non-negative integer. However, Medved
in another paper by considering some general considerations of
ensemble theory, has argued that canonical and microcanonical
corrections could not cancel out each other to result in vanishing
logarithmic prefactor in entropy[5]. Recently, Medved and Vagenas
have shown that a tunneling framework of the Hawking radiation,
effectively constrains the coefficient of logarithmic term to be
non-negative[6]. They have argued that this observation implies the
necessity for including the canonical corrections in the quantum
formulation of the black hole entropy.\\ There are other literatures
considering logarithmic corrections to the black hole entropy-area
relation(see for example[6-8]), but there is no explicit statement
about the exact value of this prefactor and specially its dependence
to spacetime dimensionality. There are several questions about the
value of this prefactor. Some of these questions are: what is the
role of the grand canonical ensemble? What is the exact role of
spin, angular momentum, {\it etc.} in the black hole thermodynamics?
Some of these questions have been discussed in literatures[4-6].
Here we won't to answer these questions. What we want to do is just
to show that from microcanonical point of view, there is a relation
between spacetime dimensionality and the existence or vanishing of
logarithmic prefactor. To elaborate our proposal, we calculate the
quantum correction of entropy-area relation using the generalized
uncertainty principle in a model universe with large extra
dimensions. We use Arkani-Hamed, Dimpolous and Dvali(ADD) model of
large extra dimensions to perform our calculations. In this model,
$n$ extra spacelike dimensions with the same radius $L$ and without
any curvature are suggested. All standard-model particles are
confined to the observable 4-dimensional brane, whereas gravitons
can access the whole d-dimensional bulk spacetime, being localized
at the brane at low energies[9].\\

\section{Thermodynamics of a d-Dimensional Schwarzschild Black Hole}
The generalized uncertainty principle is a model independent aspect
of quantum gravity proposal. It can be addressed in several
approaches to quantum gravity and a nonzero minimal observable
length naturally emerges from this proposal. Within string theory,
investigation of string collisions at Planckian energies and through
a renormalization group type analysis, the emergence of a minimal
observable distance leads to the generalized uncertainty
principle[10]
\begin{equation}
\delta x\geq \frac{\hbar}{2}\Big(\frac{1}{ \delta p} + \alpha^2
l_p^2 \frac{\delta p}{\hbar^2}\Big).
\end{equation}
Here $\alpha$ is dimensionless, positive and independent of $\delta
x$ and $\delta p$ but may in general depend on the expectation
values of $x$ and $p$.  It is on the order of unity and depends on
the details of the quantum gravity proposal. At energies much below
the Planckian energy, the extra term can be ignored and usual
uncertainty principle of Heisenberg is recovered. But at the high
energy regime, the extra term plays very important role: the
appearance of this extra term leads to the finite resolution of
spacetime points in high energy regime. In a model universe with
large extra dimensions, our GUP can be written as
\begin{equation}
\delta x_i \delta p_i \geq \frac{\hbar}{2}\Big(1+\frac{\alpha^2
L_p^2 (\delta p_i)^2}{\hbar^2}\Big),
\end{equation}
where $L_p=(\frac{\hbar G_d}{c^3})^{\frac{1}{d-2}}$ is Planck length
in extra dimensional scenario. $d$ is the total number of spacetime
dimensions, $G_d=G_{4}L^n$ where $n=d-4$ and $L$ is the extension of
$n$ extra dimensions in ADD model[11]. Now, after a brief review of
the preliminaries we calculate black hole temperature
and entropy in which follows.\\
A d-dimensional spherically symmetric black hole of mass $M$ (to
which the collider black hole will settle into before radiating) is
described by the following metric[12]
\begin{equation}
ds^2 = -\bigg(1-\frac {16\pi G_d M}{(d-2)\Omega_{d-2} c^2
r^{d-3}}\bigg) c^2 dt^2+  {\bigg( 1- \frac {16 \pi G_d
M}{(d-2)\Omega_{d-2}c^2 r^{d-3}}\bigg )}^{-1} dr^2+ r^2 d
\Omega^{2}_{d-2},
\end{equation}
where $\Omega_{d-2} = \frac{2 \pi^{\frac{d-1}{2}} }
{\Gamma(\frac{d-1}{2})}$ is the metric of the unit $S^{d-2}$ sphere.
Since Hawking radiation is a quantum gravitational phenomenon, the
emitted quanta from an evaporating black hole should be treated
within the generalized uncertainty principle framework. If we
simulate a black hole as a $(d-1)$ dimensional cube of size equal to
its Schwarzschild radius $r_s$, the uncertainty in the position of a
Hawking particle at the emission is
\begin{equation}
\delta x_i \approx r_s=\omega_{d} L_{p} m^{\frac{1}{d-3}},
\end{equation}
where
$\omega_d=\bigg(\frac{16\pi}{(d-2)\Omega_{d-2}}\bigg)^{\frac{1}{d-3}}$,\,\,
$m=\frac{M}{M_p}$ and
$M_p=(\frac{\hbar^{d-3}}{c^{d-5}G_d})^{\frac{1}{d-2}}$. Here
$\omega_d$ is a dimensionless area factor. A simple calculation
based on equation (4) leads to
\begin{equation}
\delta p_i=\frac{2\hbar\delta{x_i} \pm \sqrt{4\hbar^2 (\delta x_i)^2
-4\hbar^2 \alpha^2 L_p^2}} {2 \alpha^2 L_P^2}.
\end{equation}
To find correct limiting result, we should consider the minus sign.
The minimum position uncertainty can be obtained simply and the
result is given by
\begin{equation}
(\delta x)_{min} =\alpha L_p.
\end{equation}
Since
\begin{equation}
T_H^{GUP} = \frac{(d-3)}{4\pi}c \delta p_i,
\end{equation}
we find the following expression for the black hole temperature in a
model universe with large extra dimensions
\begin{equation}
T_H^{GUP} = \frac{(d-3)}{4\pi}\   \frac {\hbar c \delta
x_i}{\alpha^2 L_p^2}\ \bigg[1-\sqrt{1-\frac{\alpha^2 L_p^2}{\delta
x_i^2}} \bigg],
\end{equation}
where $\delta x_i$ is given by (6). This relation shows implicitly
that black hole temperature increases with spacetime dimensions,
$d$. The higher temperature yields to faster decay and less
classical properties of the black hole. As a result, in models with
large extra dimensions, black holes are hotter and
shorter-lived[13].\\
Now we look at black hole entropy. When a quantum particle with
energy $E$ and size $l$ is absorbed by a black hole, the minimum
increase in the horizon area of the black hole can be expressed as
follows
\begin{equation}
(\Delta A)_{min}\geq \frac{8 \ln2 L_p^{d-2}E l}{(d-3)\hbar c},
\end{equation}
where by $E\sim c \delta p_i$ and $l\sim \delta x_i$, can be
re-written as
\begin{equation}
(\Delta A)_{min}\geq \frac{8\ln2 L_p^{d-2}c \delta p_i \delta x_i
}{(d-3)\hbar c}.
\end{equation}
Using the value of  $\delta p_i$ from (7), we obtain
\begin{equation}
(\Delta A)_{min}\geq \frac{8\ln2 L_p^{d-4}\delta
x_i^2}{(d-3)\alpha^2}\bigg[ 1-\sqrt{1-\frac{\alpha^2 L_p^2}{\delta
x_i^2}} \ \bigg]
\end{equation}
To calculate the microcanonical entropy of a large black hole, we
put $\delta x_i\approx r_s$ and $ r_s^2 =
\Omega_{d-2}^{-\frac{2}{d-2}} A^{\frac{2}{d-2}}$ where
$\Omega_{d-2}=\frac{2\pi^{\frac{d-1}{2}}}{\Gamma(\frac{d-1}{2})}$.
If we set $(\Delta S)_{min} =b$, then we find
\begin{equation}
\frac{d S_d}{d A}\simeq  \frac{(\Delta S)_{min}}{(\Delta
A)_{min}}\simeq \frac{
\frac{(d-3)b\alpha^2\Omega_{d-2}^{\frac{2}{d-2}}}{8\ln2 L_p^{d-4}}}
{A^{\frac{2}{d-2}} \bigg[1-  \sqrt{
1+\frac{-\alpha^2L_p^2\Omega_{d-2}^{\frac{2}{d-2}}}{A^{\frac{2}{d-2}}}}\
\bigg]}
\end{equation}
For simplicity, we define $
\lambda=\frac{(d-3)b\alpha^2\Omega_{d-2}^{\frac{2}{d-2}}}{8 \ln2
L_p^{d-4}} $ and $ \eta=-\alpha^2L_p^2\Omega_{d-2}^{\frac{2}{d-2}}
$. Thus we find
\begin{equation}
\frac{d S_d}{d A}\simeq \frac{\lambda} {A^{\frac{2}{d-2}} \bigg[1-
\sqrt{ 1+\frac{\eta}{A^{\frac{2}{d-2}}}}\ \bigg]},
\end{equation}
or in integral form
\begin{equation}
S_d\simeq \lambda\int_{A_p}^A {\frac{d A} {A^{\frac{2}{d-2}}
\bigg[1- \sqrt{ 1+\frac{\eta}{A^{\frac{2}{d-2}}}}\ \bigg]} },
\end{equation}
where $ A_p=\Omega_{d-2}(\alpha L_p)^{d-2}$ is the area of event
horizon of black hole remnant. Existence of these remnants is a
consequence of minimal length conjecture[14,15]. We apply a Taylor
series expansion to obtain the following expression
\begin{equation}
S_d\simeq  -\frac{2\lambda}{\eta}\int_{A_p}^A d A \Bigg[ 1 +
\frac{\eta}{4}A^{-\frac{2}{d-2}}
-\frac{\eta^2}{16}A^{-\frac{4}{d-2}}+\frac{\eta^3}{32}
A^{-\frac{6}{d-2}}-\frac{5\eta^4}{256}A^{-\frac{8}{d-2}}
-\frac{7\eta^5}{256}A^{-\frac{10}{d-2}}+...\Bigg].
\end{equation}
In which follows, we calculate $S_d$ for some values of
$d$. Note that we choose $b=\ln2$(as one bit of information).\\
For $d=4$ (our four-dimensional brane)
$$S_4\simeq \frac{A}{4 L_p^2}-\frac{\pi\alpha^2}{4}\ln \frac{A}{4
L_p^2} +\frac{\pi^2 \alpha^4}{16}\Big(\frac{4 L_p^2}{A}\Big)
+\frac{\pi^3 \alpha^6}{16}\Big(\frac{4 L_p^2}{A}\Big)^2
+\frac{5}{3}\frac{\pi^4\alpha^8}{256}\Big(\frac{4 L_p^2}{A}\Big)^3$$
\begin{equation}
-\frac{7}{4}\frac{\pi^5\alpha^{10}}{256}\Big(\frac{4
L_p^2}{A}\Big)^{4}+...+{\cal {C}},
\end{equation}
where ${\cal {C}}$ is a constant. This relation has the standard
form of entropy-area relation as given by string theory. The matter
which should be stressed here is the fact that our calculation rules
out the possibility of a power-law correction, that is, there are no
correction terms proportional to $\Big(\frac{A}{4L_P^2}\Big)^{n}$.
On the other hand, our approaches give the value of
$c_{0}=-\frac{\pi\alpha^2}{4}$ for logarithmic prefactor. Since
$\alpha$ is a string theory parameter related to the minimal
observable length, $(\Delta x)_{min}=\alpha L_{p}$, it is a positive
quantity and therefore microcanonical prefactor is negative. If we
accept the Hod's prescription[3], we should add the positive
contribution of the canonical ensemble to find non-negative
prefactor. This result is consistent with the result of Medved and
Vagenas[6] regarding the necessity for including the canonical
corrections in the quantum formulation of the black hole entropy to
find non-negative prefactor.\\
Now we consider the effect of the extra dimensions.\\
For $d=5$, we obtain \\
$$S_5\simeq\frac{1}{2 L_p^3} \Big[ A -1.19\pi^{4/3}\alpha^2L_p^2A^{1/3}
+0.47\pi^{8/3} \alpha^4L_p^4\frac{1}{A^{1/3}} +0.13\pi^4
\alpha^6L_p^6 \frac{1}{A}
+0.07\pi^{16/3}\alpha^8L_p^8\frac{1}{A^{5/3}}$$
\begin{equation}
-0.12\pi^{20/3}\alpha^{10}L_p^{10}\frac{1}{A^{7/3}}+...+{\cal
{C}}\Big],
\end{equation}
which has no terms consisting $\ln A$ \, but contains some
extraordinary powers of area. For $d=6$, we find\\
$$S_6\simeq\frac{3}{4 L_p^4} \Big[ A -0.81\pi\alpha^2L_p^2A^{1/2}
-0.17\pi^{2} \alpha^4L_p^4\ln\Big(\frac{A}{4L_{p}^{2}}\Big)
+0.27\pi^3 \alpha^6L_p^6 \frac{1}{A^{1/2}}
+0.035\pi^{4}\alpha^8L_p^6\Big(\frac{4L_{p}^{2}}{A}\Big)$$
\begin{equation}
-0.21\pi^{5}\alpha^{10}L_p^{10}\frac{1}{A^{3/2}}+...+{\cal
{C}}\Big],
\end{equation}
with an explicit logarithmic correction term and some extraordinary
area dependent terms. For $d=7$, we find\\
$$S_7\simeq\frac{1}{ L_p^5} \Big[ A -0.42\pi^{6/5}\alpha^2L_p^2A^{3/5}
-0.31\pi^{12/5} \alpha^4L_p^4A^{1/5} +0.16\pi^{18/5} \alpha^6L_p^6
\frac{1}{A^{1/5}}
+0.03\pi^{24/5}\alpha^8L_p^8\frac{1}{A^{3/5}}$$
\begin{equation}-
0.03\pi^{6}\alpha^{10}L_p^{10}\frac{1}{A}+...+{\cal {C}}\Big],
\end{equation}
without any logarithmic correction term. For $d=8$, our calculations gives\\
$$S_8\simeq\frac{5}{4 L_p^6} \Big[ A -0.38\pi\alpha^2L_p^2A^{2/3}
-0.19\pi^{2} \alpha^4L_p^4A^{1/3} -0.03\pi^{3} \alpha^6L_p^6
\ln\frac{ A}{4L_{p}^{2}}
+0.06\pi^{4}\alpha^8L_p^8\frac{1}{A^{1/3}}$$
\begin{equation}
-0.05\pi^{5}\alpha^{10}L_p^{10}\frac{1}{A^{2/3}}+...+{\cal
{C}}\Big],
\end{equation}
with an explicit logarithmic correction term. For $d=9$, we find\\
$$S_9\simeq\frac{3}{2 L_p^7} \Big[ A
-0.26\pi^{8/7}\alpha^2L_p^2A^{5/7} -0.08\pi^{16/7}
\alpha^4L_p^4A^{3/7} -0.08\pi^{24/7} \alpha^6L_p^6 A^{1/7}
+0.04\pi^{32/7}\alpha^8L_p^8\frac{1}{A^{1/7}}$$
\begin{equation}-
0.01\pi^{40/7}\alpha^{10}L_p^{10}\frac{1}{A^{3/7}}+...+{\cal
{C}}\Big],
\end{equation}
without any logarithmic correction term. And finally, for $d=10$, we
find
$$S_{10}\simeq\frac{7}{4 L_p^8} \Big[ A -0.25\pi\alpha^2L_p^2A^{3/4}
-0.07\pi^{2} \alpha^4L_p^4A^{1/2} -0.05\pi^{3} \alpha^6L_p^6 A^{1/4}
-0.01\pi^{4}\alpha^8L_p^8\ln \frac{A}{4L_{p}^{2}}$$
\begin{equation}
-0.02\pi^{5}\alpha^{10}L_p^{10}\frac{1}{A^{1/4}}+...+{\cal
{C}}\Big],
\end{equation}
which has a logarithmic correction term. Note that $L_{p}$ and
${\cal {C}}$ have different values in different spacetime
dimensions. From these relations, one can deduce the following
conclusions
\begin{itemize}
\item
In four dimensional spacetime(our brane), our approach gives the
standard prescription of string theory as given by the relation (2)
or
\begin{equation}
S_{4}=\frac{A}{4L_p^2}+ c_0 \ln(\frac{A}{4L_p^2})+\sum_{n=1}^\infty
c_n (\frac{A}{4L_p^2})^{-n} + {\cal {C}}.
\end{equation}
There is a logarithmic correction term with exact value of
prefactor, $c_{0}=-\frac{\pi\alpha^2}{4}$. Our approach rules out
the possibility of power-law expansion of correction terms in four
dimensions.
\item
Comparison between the functional form of entropy in different
spacetime dimensionalities, shows that the microcanonical
logarithmic prefactor appears only in spacetimes with even number of
dimensions. For spacetimes with odd dimensionality, there is no
microcanonical logarithmic prefactor.
\item
For $d>4$, we observe that some unusual powers of area have been
appeared in entropy formula. These unusual terms are not consistent
with general prescription as is described by relation (2). One may
argue that entropy-area relation in extra dimensional scenarios do
not obeys the prescription provided by (2). Even it is possible to
have a power-law correction of entropy-area relation in extra
dimensions. If we insist on the validity of the prescription (2), we
should omit unusual terms in entropy-area relations for $d>4$. This
procedure leads to severe constraints on the functional form of
modified dispersion relations(see [2] and references therein). But
there is no obvious reason for omitting these terms in extra
dimensional scenarios. The only statement that one can say is that
the higher dimensional scenarios lead to entropy-area relations
which have considerable departure from their four-dimensional
counterpart.
\item
The existence or vanishing of logarithmic prefactor not only depends
on the spacetime dimensionality but also depends on the statistical
ensemble used in the course of the calculations. We have shown that
this prefactor in the case of microcanonical ensemble, exists only
when spacetime has an even number of dimensions and is negative.
Comparison between our finding and the results of Hod and Medved,
shows that to have a non-negative prefactor, the contribution of the
canonical ensemble should be positive. But this is not the complete
argument of the issue since contributions of grand-canonic ensemble
and spin effects should be taken into account. In this paper we have
shown the explicit role of the microcanonical ensemble.
\end{itemize}

\end{document}